\documentclass[prb,preprint]{revtex4-1} 

\usepackage{amsmath}  
\usepackage{amsfonts} 
\usepackage{graphicx} 

\begin{document}

\title{3D Printing an External Cavity Diode Laser Housing}

\author{E. Brekke}
\email{erik.brekke@snc.edu}
\author{T. Bennett}%
\affiliation{Department of Physics, St. Norbert College, De Pere, WI 54115
}%

\author{H. Rook}
\affiliation{
 Department of Physics, Carleton College, Northfield, MN 55057
}%

\author{E.L. Hazlett}
\affiliation{
 Department of Physics, St. Olaf College, Northfield, MN 55057
}%
\date{\today}

\begin{abstract}
The ability to control the frequency of an external-cavity diode laser (ECDL) is an essential component for undergraduate laboratories and atomic physics research.  Typically the housing for the ECDL's diffraction grating and piezoelectric transducer is either purchased commercially or machined from metal.  Here, we present an alternative to these commonly used options that utilizes 3D printing, a tool available in many physics departments.  We characterize the performance of our ECDL system using atomic spectroscopy and self-heterodyne interferometry and show that it is sufficient for use in undergraduate spectroscopy experiments and a number of research applications where extremely narrow laser linewidths are not necessary.  The performance and affordability of 3D-printed designs make them an appealing option for future use.  
\end{abstract}

\maketitle


\section{\label{sec:intro}Introduction}
Laser diodes are now commonplace in both undergraduate experiments and atomic research.  While the diodes themselves are cheap and easy to 
obtain, stable and controllable operation of these diodes requires implementing of an optical feedback method.  Most commonly, this needed feedback is accomplished through the use of a diffraction grating in the Littrow configuration. In this method, the diffraction grating forms an external cavity of the laser and, by adjusting the angle of the grating, the frequency of the resonant light in this external cavity is controlled. By integrating a piezoelectric transducer in this system that provides feedback for the adjustment of the grating angle, a frequency-stabilized diode is attained.  \cite{Fleming81}  The laser housing for such an extended-cavity diode laser (ECDL) is an assembly that supports and allows for the adjustment of the essential components. Most commonly, this assembly is purchased commercially or is constructed from machined aluminum components.\cite{Wieman92, Boshier98} The latter approach works well, but requires access to machining tools and expertise, which are often barriers for implementation. 

The method presented here breaks down these barriers by using 3D printing technology. With recent significant decreases in
the cost of 3D printers and materials as well as access to computer-aided design (CAD) programs, 3D printing is now common in most educational and industrial environments. Hence, these fabrication technologies have found a number of uses in physics instruction and research. \cite{Virgin:18, Su19, Virgin2017, Heinrich:2020} 

In this paper, we will show that modern 3D printing techniques provide a viable alternative for the creation of necessary components for an ECDL system. Here, we present a laser assembly that enables optical feedback with the Littrow method. As shown in Fig. \ref{fig:assembly}, our assembly is constructed from 3D printed materials and simple low-cost commercial components.   We will show that this design has a large mode-hop-free tuning range, a sufficient short-term stability, and the adaptability necessary to make it an excellent option for undergraduate laboratories and atomic research.

\section{\label{sec:design}Design}

The essential design elements of an ECDL assembly provide support for a diffraction grating, ability to adjust the vertical and horizontal location of the optical feedback beam, and positioning of a piezoelectric transducer to enable scanning of the laser frequency. The goal of the design presented here is to incorporate these elements into a 3D printed version, which requires no additional machining. For our system, the components that cannot be 3D printed are easily available, ensuring the broad accessibility of the design.  For a complete materials list, CAD files, and detailed assembly instructions see the supplemental information. \cite{3Dfiles} 
We used a TazBot Lulz 6 printer, which is a Fused Filament Fabrication printer, with PLA (Polylactic Acid) filament.\cite{lulz} 

\begin{figure}[htbp]
\centering
\includegraphics[width=15.5cm]{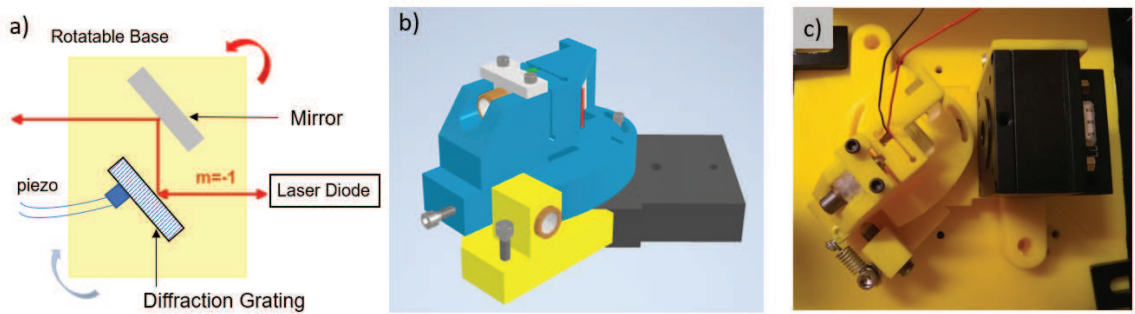}
\caption {a) The basic operation of a Littrow ECDL showing optical feedback to the laser diode from the $m=-1$ diffracted order. The blue/red arrows indicate the direction of rotation needed to increase/decrease the frequency of the feedback.   b)  A CAD view of the 3D laser housing showing the main components.  The base in dark grey, the feedback housing in blue, and coarse adjustment housing in yellow. The grating is in red.  The laser diode is mounted to the base via the mounting hole on the right  c) Photo of an assembled laser setup. The base in the CAD drawing is designed to be mounted on pedestals, but was modified for optical breadboard mounting for the picture. }
\label{fig:assembly}
\end{figure}

Figure \ref{fig:assembly} depicts the principle of optical feedback for an ECDL and the key components of the design. The first-order diffraction ($m=-1$) of the emitted laser light provides the feedback that controls the frequency of the laser operation at a desired frequency. Light of the same order, but of higher (lower) frequency are deflected above (below) the resonant $m=-1$ diffracted order. The stabilized $m=0$ order light is then reflected off a mirror that rotates with the grating to maintain alignment of the light.  These optics are mounted on a rotatable base to control the frequency in a coarse manner as indicated by the color of the arrows. A piezoelectric transducer provides fine adjustment of the grating for electronic control of the optical feedback. 


In our assembly, the diffraction grating is attached to the 3D housing base via super-glue or epoxy. An 1800 lines/mm grating was used, giving a 45-degree angle of operation at 780 nm. The coarse angle adjustment is done with a fine-thread set screw and spring, which can be locked into place with two 4-40 set screws.  A reflection mirror is used to ensure constant output steering.\cite{HawthornOutput}  It is important to choose a mirror with a thin profile (less than 3 mm) in order to prevent multiple reflections off of the grating.  If the experiment is not sensitive to output steering, or if the laser will not be tuned extensively, the mirror can be eliminated completely. This modification can be done by adjusting the CAD file or by removing the arm in post-print.

The rotation of the feedback housing has a large travel range, but not all of this range is usable due to multiple reflections from the steering mirror and the limited distance of travel of the coarse adjustment screw.  Both of these issues can be easily modified in the CAD files to tailor to the base for a specific diode. For the version shown here, the coarse angle scan range is limited to 42-48 degrees with respect to the normal of the grating. Taking these angles into account, the estimated compatible wavelength ranges are from to 743--825 nm and 1115--1238 nm range with 1800 lines/mm and 1200 lines/mm grating, respectively.  The edges of these ranges are dependent on the grating being mounted flush against the base. The angle of the grating mount surface can be adjusted to accommodate wavelength of other diodes. We have implemented this change successfully to tune a laser near 1010 nm.  Elimination of the steering mirror expands the tuning range, at a sacrifice to the stability of the output-beam direction. 

Once the coarse angle alignment is set, the vertical angle is adjusted with an additional fine-thread screw. Here the natural elasticity of the 3D printed housing is used to maintain the alignment instead of a spring.  A 5-degree vertical pitch is built into the grating mounting to ensure that proper vertical alignment occurs in the middle of an adjustable range. 

A piezo transducer is used for the fine angle adjustment, counterbalanced with a built in elastic cantilever, similar to the vertical alignment system. For ABS (Acronylite Butadiene Styrene) and PLA, the surface hardness is large enough that the piezo can tune the wavelength over several GHz with minimal hysteresis. 

Due to shrinkage and different slice settings, the housing for the piezo transducer may be larger than the piezo itself. This problem can be alleviated by either printing test pieces or through the insertion of  small metal shims that are easily available at any hardware store or online. It is important to note that although the location of the coarse horizontal adjustment assembly is not positioned for optimal scanning range, the piezo scan pivot provides synchronous cavity mode and feedback wavelength scanning. \cite{McNicholl85}

There are various ways to secure the laser to an optical table. The base design included in the supplemental material is shown in Fig. \ref{fig:assembly}(b). This design includes a set socket for a 8-32 set screw that is compatible with an optical post or pedestal, or table clamps can be used to secure the base to an optical table or breadboard. The base can also be easily modified to allow for a direct mounting to a 1-inch grid optical table, as shown in Fig. \ref{fig:assembly}(c).

It is expected that the exact performance of the laser will depend on the material used and the filling geometry of the print.  Here PLA was shown to be effective, with the tetrahedral infill seeming to provide extra stability.  Fills between 20\% and 100\% were used, with the balance of flexibility and stability being especially important for the vertical adjustment.  The print parameters could be further investigated to determine the ideal characteristics for particular applications.

\section{\label{sec:materials} Materials and Cost}
A great benefit of 3D printing is the reduction of the cost barrier to scientific investigations such as the Foldscope \cite{foldscope} and hand-powered blood centrifuge.\cite{centrifuge} In this spirit, one of the goals of our design was to keep the cost as low as possible.  We estimate a cost of less than
 \$20 for 3D printing the necessary components.  
 
\begin{table}[b]
\caption{\label{tab:table}
The parts required for construction of the 3D printed ECDL system.  Possible retailers with part numbers are supplied, but equivalent pieces can be found from a wide number of retailers.}
\begin{ruledtabular}
\begin{tabular}{cccccccc}
 Part&Specs&Supplier and part number&
 Price (\textdollar)\\
\hline
Diffraction Grating&1800 lines/mm 12.7mm square&Dynasil G1800R240CEAS & 57 \\
Mirror& square 12.7mm, $<$3mm thick& Thorlabs ME05S-P01 & 17.75 \\
Laser Diode& 780 nm & BlueSky VPSL-785-025-Q-5-B & 78 \\
Diode Mount&  & Thorlabs LDM21 &364.67 \\
Piezo& 150 V Piezo, 5 mm thick & Thorlabs PK4DLP2 &33.28 \\
Adjustment Screws& 2x 1/4"-100 fine adjustment & Thorlabs F25US075 &10.92 \\
Adjustment Bushings& 2x 1/4"-100 fine adjustment & Thorlabs F25USN1P & 11.48 \\
Hardware& 8 x 4-40 screws 3/8" & various & 5 \\
Spring& short tension & various & 5 \\
Adhesive& Epoxy or Superglue & various & 5 \\
\hline
&  & Total & 588  \\
\end{tabular}
\end{ruledtabular}
\end{table}

 Table \ref{tab:table} lists the components needed for the full assembly.  These components comprise the main optics and optomechanics for the ECDL and put the cost of this project at less than \$600, which is an order of magnitude less than the cost of a commercial ECDL. To fully utilize the system, stabilization of the laser-diode temperature and drive current is required.

In our setup, we used a commercial diode laser current driver (Thorlabs LDC202C)  and temperature controller (Thorlabs TED200C) costing approximately \$1000 each.  This significantly increases the cost of the laser system. By integrating our ECDL with home-built laser diode current drivers,\cite{DufreeDriver,LibbrechtDriver} piezo drivers,\cite{PiezoDriver} and temperature controllers,\cite{BaroneTemp} the cost for a complete atomic spectroscopy can be significantly reduced. One can also build a laser diode mount with integrated thermoelectric cooling ability to replace the Thorlabs LDM21 mount and its cooling ability. This mount would require either the laser diode to have the same pointing as the LMD21 mount or for the CAD files to be adjusted.

This setup will not replace the commercial ECDL in all research applications, but it can replace it in simple spectroscopy experiments and experimental components.  In addition, this design is extremely beneficial for upper division lab courses, giving students access to laser and spectroscopy experiments. The cost advantages and ease of use make this system an excellent option for a number of atomic experiments.

\section{\label{sec:testing}Diagnostics and Characteristics}

 In order to demonstrate the capabilities and characterize our 3D printed system, we tested its performance using spectroscopy and self-heterodyne analysis. For our first spectroscopy experiment, we performed saturated absorption spectroscopy on a rubidium vapor.\cite{Wieman92, Preston96}  This setup was used with the 3D printed system to examine the laser behavior and determine the mode-hop-free tuning range and frequency resolution.  Using the horizontal gross adjustment, the laser can easily be scanned over several nanometers and tuned to the region of atomic transitions.

\begin{figure}
\centering
\includegraphics[width=11.2cm]{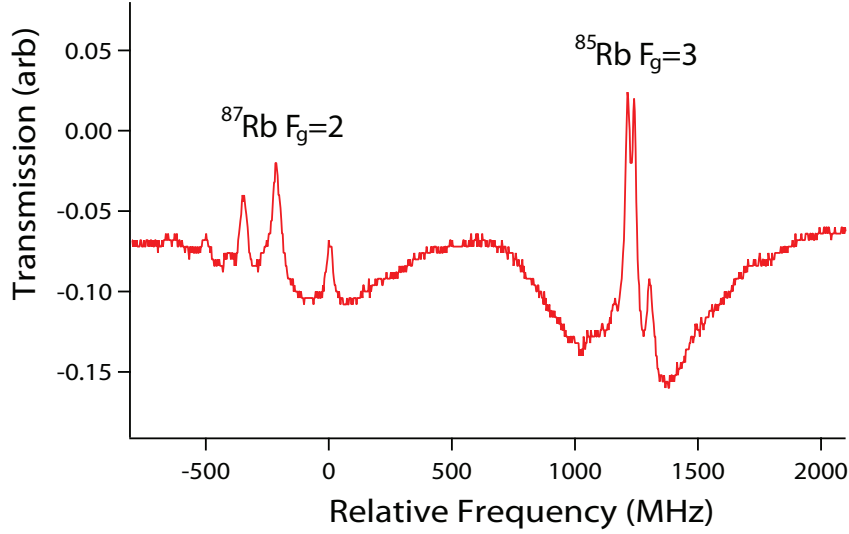}
\caption {A scan of the laser frequency over the rubidium isotope lines.  This shows a scanning range of over 2 GHz.  The system can also resolve the crossover peaks in $^{85}$Rb, with a separation of 32 MHz.}
\label{fig:SApeaks}
\end{figure}

The observed transmission spectrum for the rubidium $5S_{1/2}$ to $5P_{3/2}$ transition for natural isotope abundance is shown in Fig. \ref{fig:SApeaks}.  While scanning the laser piezo, the 3D printed system works well for displaying the key features of saturated absorption.  These data demonstrate a mode-hop-free tuning range of over 2 GHz and the ability to resolve peaks separated by 30 MHz.  

\begin{figure}[htbp]
\centering
\includegraphics[width=11.2cm]{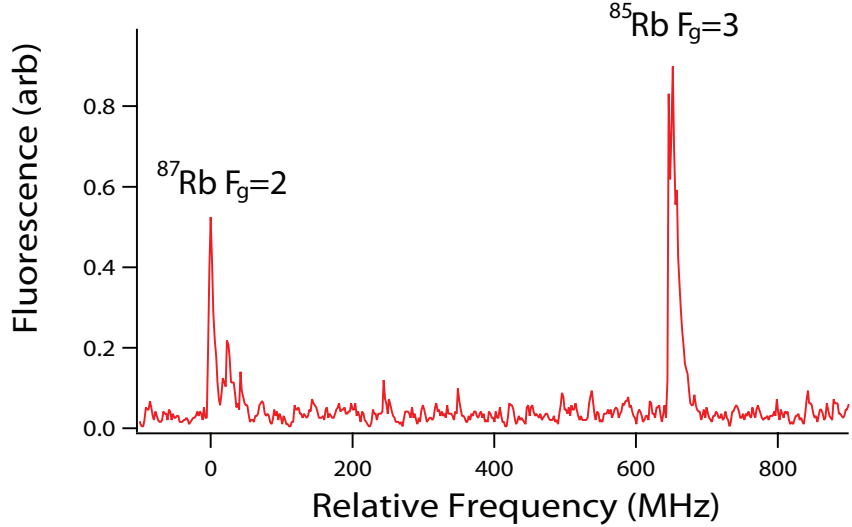}
\caption {Two-photon spectroscopy showing 420 nm fluorescence as a function of the laser frequency as scanned over the $5S_{1/2}$ to $5D_{5/2}$ transition.  Note the laser frequency separation is one half the ground state hyperfine splitting for the two photon transition.}
\label{fig:twophoton}
\end{figure}

To gain further insight into the properties of the 3D printed system, it was also used at 778 nm for two photon spectroscopy of the rubidium $5S_{1/2}$ to $5D_{5/2}$ transition. \cite{Olson06}  The laser was found to scan without a mode-hop across hyperfine transitions for the different isotopes.  A scan showing the peaks for the $^{87}$Rb $F=2$ and $^{85}$Rb $F=3$ transitions is shown in Fig. \ref{fig:twophoton}.  Due to the small hyperfine splitting in the $5D$ state, this scan demonstrates the ability to resolve peaks whose excitation frequency differs by only 8 MHz for the $^{87}$Rb, though the peaks in $^{85}$Rb at 3 MHz to 5 MHz separations were not resolved.  

\begin{figure}[htbp]
\centering
\includegraphics[width=11.2cm]{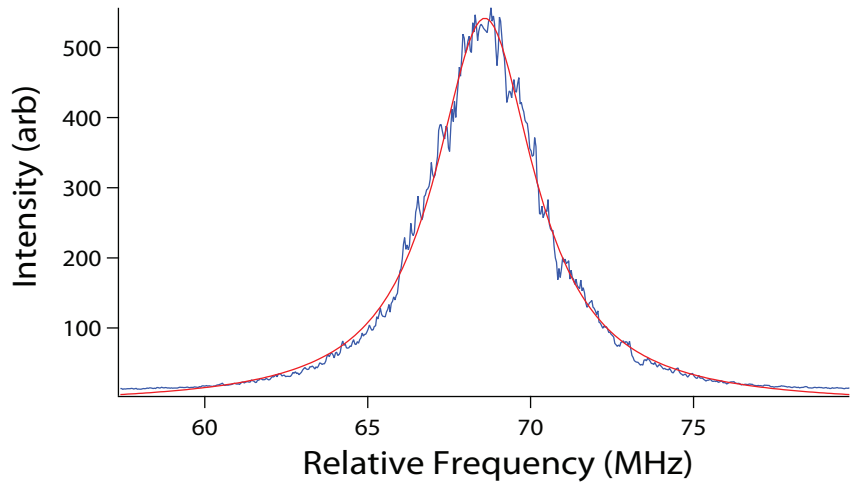}
\caption {The spectrum for the auto-correlation observed using the self-heterodyne technique with a 1 second integration time.  This gave a laser linewidth of $1.7 \pm 0.3$ MHz.  }
\label{fig:width}
\end{figure}

The system design was further tested using a self-heterodyne method \cite{Nakayama80, McGrath86} in order to examine the short-term linewidth. The laser was split, with half the beam going through an 11 km delay line before recombination, in the traditional self-heterodyne setup.  An AOM was used in one of the paths to shift the center frequency by 68.5 MHz for easier observation in a spectrum analyzer.  
The spectrum was averaged over a 1 second time interval, with the resulting spectrum shown in Fig. \ref{fig:width}.  The linewidth of the laser was measured to be $1.7 \pm 0.3$ MHz.  This linewidth is sufficient for use in undergraduate spectroscopy experiments and a number of research applications where extremely narrow linewidths are not necessary.  It is likely that this linewidth could be further reduced through better protection of the system from vibration and air currents.

As most of the readily available 3D-printing materials are various types of plastics, there is some concern about creep and degradation of the elastic properties of the system. In our investigation, we have found that the integrity of the mount was maintained over the course of weeks to months with only a minimal amount of adjustment to the system needed to keep it on an atomic resonance line, signifying that creep and degradation is not a hindrance of performance over that time scale.  While this design has not been investigated with laser locking, we expect that it should allow for standard locking mechanisms.  

\section{\label{sec:futhure}Future directions}

The laser housing design shown here has demonstrated tunability, stability, and adaptability for a large number of atomic applications.  For those without access to metal machining, this 3D-printed design brings an ECDL to an achievable price point.  While those with machining ability may find cheaper and easier alternatives, 3D printing offers the ability for quick prototype turnaround and is widely available. As the state of the art in 3D printing becomes more available, this model could be further adapted.  The current form of this design has demonstrated the ability to perform spectroscopy, and there are still advances that can be made.  The print material and fill can be further optimized, as well as implementation on resin-based or metal printers. If further stabilization is necessary, adapting the design to allow for an airtight box surrounding the system, along with electrical connections that do not exert tension on the laser mount, would be helpful.  

Additional improvements include using a filament with high thermal conductivity to create a 3D-printed housing for the laser diode and thermoelectric cooler, eliminating the need for the commercial Thorlabs diode mount.  Though this modification would add additional heat to the 3D printed material which may cause deformation, with careful planning of the geometry of the print and multiple print materials these effects could be suppressed. \cite{XU201854} In addition, different base and support print geometries can be scaled to suppress external vibrations via phononic band gaps. \cite{Matlack8386}

\begin{acknowledgments}
We wish to acknowledge the contributions of Brandon Nelson to initial designs of the ECDL, as well as Logan Hennes for testing adaptations to other wavelengths.
\end{acknowledgments}






\begin{thebibliography}{99}
\bibitem [{lul()}]{lulz}%
  \BibitemOpen
  \href@noop {} {}\bibinfo {note} {The adjusted range of the final product
  allows for flexibility in the precision of the 3D printer used. Our choice
  was made due to availability. Any printer with a 4-inch by 4-inch bed could
  be used.}\BibitemShut {Stop}%
\end{thebibliography}
\end{document}